\documentclass{article}
\usepackage{spconf,amsmath,graphicx}
\usepackage{booktabs}

\usepackage{amssymb}
\usepackage{amsfonts}

\usepackage[
backend=biber,
style=ieee,
maxbibnames=5,
maxcitenames=5,
doi=false,isbn=false,url=false,eprint=false
]{biblatex} 

\defbibheading{bibliography}[\refname]{}

\DeclareSourcemap{
	\maps[datatype=bibtex, overwrite=true]{
		\map{
			\step[fieldsource=booktitle,
			match=\regexp{.*Interspeech.*},
			replace={Proc. INTERSPEECH}]
			\step[fieldsource=journal,
			match=\regexp{.*INTERSPEECH.*},
			replace={Proc. INTERSPEECH}]
			\step[fieldsource=booktitle,
			match=\regexp{.*ICASSP.*},
			replace={Proc. ICASSP}]
			\step[fieldsource=booktitle,
			match=\regexp{.*International.*Conference.*on.*Acoustics,.*Speech.*and.*Signal.*Processing.*},
			replace={Proc. ICASSP}]
			\step[fieldsource=booktitle,
			match=\regexp{.*International.*Conference.*on.*Learning.*Representations.*},
			replace={Proc. ICLR}]
			\step[fieldsource=booktitle,
			match=\regexp{.*International.*Conference.*on.*Machine.*Learning.*},
			replace={Proc. ICML}]
			\step[fieldsource=booktitle,
			match=\regexp{.*Automatic.*Speech.*Recognition.*and.*Understanding.*},
			replace={Proc. ASRU}]
			\step[fieldsource=booktitle,
			match=\regexp{.*Spoken.*Language.*Technology.*},
			replace={Proc. SLT}]
			\step[fieldsource=booktitle,
			match=\regexp{.*Speech.*Synthesis.*Workshop.*},
			replace={Proc. SSW}]
			\step[fieldsource=booktitle,
			match=\regexp{.*workshop.*on.*speech.*synthesis.*},
			replace={Proc. SSW}]
			\step[fieldsource=booktitle,
			match=\regexp{.*Advances.*in.*neural.*information.*processing.*},
			replace={Proc. NeurIPS}]
			\step[fieldsource=journal,
			match=\regexp{.*Advances.*in.*Neural.*Information.*Processing.*},
			replace={Proc. NeurIPS}]
			\step[fieldsource=booktitle,
			match=\regexp{.*Workshop.*on.* Applications.* of.* Signal.*Processing.*to.*Audio.*and.*Acoustics.*},
			replace={Proc. WASPAA}]
			\step[fieldsource=booktitle,
			match=\regexp{.*International.*Symposium.*on.*Robot.*and.*Human.*Interactive Communication.*},
			replace={Proc. RO-MAN}]
			\step[fieldsource=booktitle,
			match=\regexp{.*International.*Conference.*on.*Language.*Resource.*and.*Evaluation.*},
			replace={Proc. LREC}]
			\step[fieldsource=publisher,
			match=\regexp{.+},
			replace={{}}]
			\step[fieldsource=month,
			match=\regexp{.+},
			replace={{}}]
			\step[fieldsource=location,
			match=\regexp{.+},
			replace={{}}]
			\step[fieldsource=address,
			match=\regexp{.+},
			replace={{}}]
			\step[fieldsource=organization,
			match=\regexp{.+},
			replace={{}}]
			\step[fieldsource=booktitle,
			match=\regexp{.*International.*Conference.*on.*Multimedia.*and.*Expo.*},
			replace={Proc. ICME}]
			\step[fieldsource=booktitle,
			match=\regexp{.*Conference.*on.*Computer.*Vision.*and.*Pattern.*Recognition.*},
			replace={Proc. CVPR}]
			\step[fieldsource=booktitle,
			match=\regexp{.*European.*Conference.*on.*Computer.*Vision.*},
			replace={Proc. ECCV}]
			\step[fieldsource=booktitle,
			match=\regexp{.*AAAI.*Conference.*on.*Artificial.*Intelligence.*},
			replace={Proc. AAAI}]
			\step[fieldsource=booktitle,
			match=\regexp{.*North.*American.*Chapter.*of.*the.*Association.*for.*Computational.*Linguistics.*Human.*Language.*Technologies.*},
			replace={Proc. NAACL-HLT}]
		}
	}
}

\def\x{{\mathbf x}}
\def\y{{\mathbf y}}
\def\u{{\mathbf u}}

\def\A{{\mathbf A}}
\def\B{{\mathbf B}}
\def\C{{\mathbf C}}
\def\D{{\mathbf D}}
\def\Ab{\overline{{\mathbf A}}}
\def\Bb{\overline{{\mathbf B}}}
\def\Cb{\overline{{\mathbf C}}}
\def\Db{\overline{{\mathbf D}}}
\def\K{{\mathbf K}}
\def\I{{\mathbf I}}

\title{Structured state space decoder for speech recognition and synthesis}

\name{Koichi Miyazaki, Masato Murata, Tomoki Koriyama}
\address{
CyberAgent, Inc. \\
\{miyazaki\_koichi\_xa, murata\_masato, koriyama\_tomoki\}@cyberagent.co.jp
}
\begin{document}
\ninept
\maketitle
\begin{abstract}
Automatic speech recognition (ASR) systems developed in recent years have shown promising results with self-attention models (\textit{e.g.}, Transformer and Conformer), which are replacing conventional recurrent neural networks.
Meanwhile, a structured state space model (S4) has been recently proposed, producing promising results for various long-sequence modeling tasks, including raw speech classification.
The S4 model can be trained in parallel, same as the Transformer model.
In this study, we applied S4 as a decoder for ASR and text-to-speech (TTS) tasks by comparing it with the Transformer decoder.
For the ASR task, our experimental results demonstrate that the proposed model achieves a competitive word error rate (WER) of 1.88\%/4.25\% on LibriSpeech test-clean/test-other set and a character error rate (CER) of 3.80\%/2.63\%/2.98\% on the CSJ eval1/eval2/eval3 set.
Furthermore, the proposed model is more robust than the standard Transformer model, particularly for long-form speech on both the datasets.
For the TTS task, the proposed method outperforms the Transformer baseline.

\end{abstract}


\section{Introduction}
\label{sec:intro}
End-to-end automatic speech recognition (E2E-ASR) has become popular because of its simple training process and high recognition accuracy~\cite{graves2014towards,chan2016listen,watanabe2017hybrid}.
Generally, E2E-ASR is based on a sequence-to-sequence framework consisting of an encoder that processes acoustic features and a decoder that outputs linguistic information such as phonemes and characters.
Many recent E2E-ASR systems are developed based on a Transformer \cite{vaswani2017attention} that uses self-attention layers in the encoder and decoder~\cite{dong2018speech}.
As a noticeable example, Gulati \textit{et al.} proposed Conformer~\cite{Gulati2020-conformer} that incorporates a convolutional neural network (CNN) into the encoder to explicitly capture local features and achieved state-of-the-art performance.

Transformer-based models have also achieved promising results in the speech field for various tasks such as text-to-speech (TTS), speech translations, and speech separation~\cite{guo2021recent}.
The advantage of self-attention is that it has a flexible function based on the similarity matrix that can capture global characteristics.
However, self-attention has computational complexity problem for long sequences because both the computation time and memory usage are quadratic in the sequence length.
To address these issues, customized attention layers are proposed to reduce attention computational complexity~\cite{katharopoulos2020transformers-linear, zaheer2020big-bird}.

Furthermore, as self-attention itself has no positional information for handling token order, such information must be explicitly provided as additional information.
The position information in the vanilla Transformer is \textit{positional encoding} in which the absolute position information is represented by a set of sinusoidal curves \cite{vaswani2017attention}.
This absolute position information causes overfitting problems to the sequence lengths in training data, which results in the performance degradation for long-form sequences.
Although one solution involves the use of relative position information \cite{shaw-etal-2018-self, dai-etal-2019-transformer-xl, wang2020position}, it tends to increase the computational complexity and cause incompatibility with the customized attention layers.

An alternative approach is to use other flexible function layers that can explicitly capture the position information.
In this study, we focus on a structured state space model (S4)~\cite{gu2022efficiently,gu2021combining-lssl} in which the relationship among latent state spaces are represented by linear transformation.
S4 has the characteristics of recurrent neural network (RNN) that can be applied to autoregressive generation without masking and save memory usage during inference unlike Transformer. S4 also has the property of CNN that enables parallel computing.
S4 solves the problems of the Transformer regarding the computational complexity and position information, and it outperformed in several tasks~\cite{gu2022efficiently}.
It is also reported that S4 is effective in autoregressive inference such as waveform generation~\cite{goel2022sashimi}, language modeling~\cite{gu2022efficiently}, and time-series forecasting~\cite{gu2022efficiently}.

In this paper, we propose an S4-based decoder for sequence-to-sequence speech model.
Specifically, we replace the self-attention-based decoder of Conformer ASR by the S4-based one.
We expect that the performance of S4 on autoregressive inference can also be seen in E2E-ASR framework.
Furthermore, we evaluate the performance of S4 decoder in autoregressive TTS.
Our experimental evaluation results on ASR tasks demonstrate that our proposed model achieves a competitive recognition accuracy on the datasets of LibriSpeech and Corpus of Spontaneous Japanese (CSJ) compared with the Transformer and Conformer models.
Furthermore, we show that our proposed model is more robust than the standard Transformer ASR, particularly for long-form speech.
We also show that TTS with the S4 decoder enhances the naturalness of synthetic speech compared with Transformer-TTS.

\section{S4 decoder for sequence-to-sequence speech models}
\label{sec:method}
\subsection{Structured state space model (S4)}
A structured state space model (S4) is based on linear state space layer (LSSL) \cite{gu2021combining-lssl}. Let $\u(t) \in \mathbb{R}^{D_\mathrm{in}}$ and $\y(t) \in \mathbb{R}^{D_\mathrm{out}}$ be the input and output continuous-time sequences and $\x(t) \in \mathbb{R}^N$ be a latent space sequence; then the output sequence can be obtained with the following equation:
\begin{align}
    \frac{d \x(t)}{d t} &= \A \x(t) + \B \u(t), \\
    \y(t) &= \C \x(t) + \D \u(t).
\end{align}
Using bilinear discretization, the LSSL for the discrete-time sequence sampled with a trainable step size $\Delta$ can be represented as
\begin{align}
        \x_k &= \Ab \x_{k-1} + \Bb \u_k \label{discrete_ssm}, \\
        \y_k &= \Cb \x_k + \Db \u_k, \\
        \Ab &= (\I-\Delta / 2 \cdot \A)^{-1} (\I+\Delta / 2 \cdot \A), \\
        \Bb &= (\I-\Delta / 2 \cdot \A)^{-1} \Delta\B, \\
        \Cb &= \C, \quad \Db = \D.
\end{align}
As indicated by the equations, the LSSL has the characteristics of an RNN.

The architecture of LSSL can be considered as a CNN.
By setting $\x_{-1} = \mathbf{0}$ and unrolling Eq.~(\ref{discrete_ssm}), we obtain
\begin{align}
    \y_k &= \Cb \Ab^{k}\Bb \u_0
                + \cdots + \Cb \Ab\Bb \u_{k-1}
                + \Cb\Bb \u_k + \Db \u_k.
\end{align}
Hence, the output $\y_k$ was calculated by a convolution kernel $\overline{\K}$ as follows:
\begin{align}
    \y_k &= (\overline{\K} \ast (\u_{k-L}, \dots, \u_k)) + \Db \u_k, \\
    \overline{\K} &= (\Cb \Bb ,\Cb\Ab\Bb,\dots, \Cb\Ab^{L-1}\Bb),
\end{align}
where $L$ is the kernel size.
Thus, we can utilize the fast parallel computation with GPU by calculating the convolution kernel in advance.

The LSSL has problems with regard to the instability of the state space sequence and the computational complexity in the kernel calculation.
Thus, a structured state space model called \textit{S4} is proposed to overcome the problems by restricting the state matrix $\A$ as a normal plus low-rank (NPLR) matrix \cite{gu2022efficiently}\footnote{We referred to the updated version available at \url{https://arxiv.org/abs/2111.00396v3}}, which is parameterized by
$\A = \operatorname{diag}[{\boldsymbol{\lambda}}] - \mathbf{p} \mathbf{p}^{*}$ and $\boldsymbol{\lambda} \in \mathbb{C}^{N}$,
$\mathbf{p} \in \mathbb{C}^{N}$.
We refer to this LSSL layer as the \textit{S4 layer}.
This parameterization is a structured representation that allows faster computation based on efficient computation algorithms.

\subsection{S4 decoder and its application to sequence-to-sequence speech modeling}
\begin{figure}
    \centering
    \includegraphics[clip,width=1.0\columnwidth]{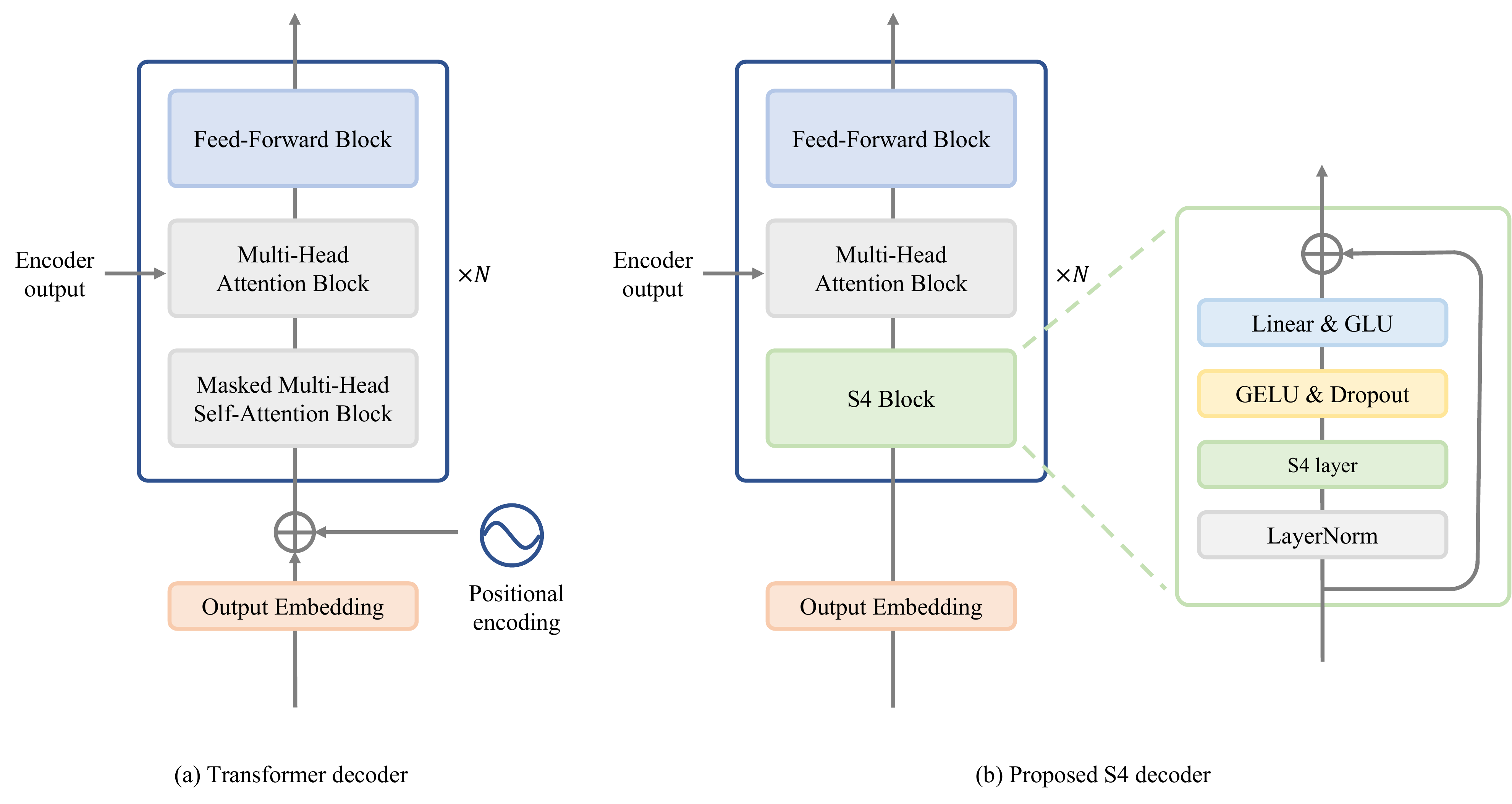}
    \caption{\it{Overview of our proposed S4 decoder architecture. (a) is an illustration of the Transformer decoder and (b) is an illustration of our proposed S4 decoder.
Compared to the Transformer decoder, the masked multi-head self-attention block is replaced by the S4 block and positional encoding is removed.}}
    \label{fig:proposed}
\end{figure}

We propose an S4 decoder that inherits the Transformer decoder architecture~\cite{vaswani2017attention}, which consists of stacks of feed-forward block, multi-head attention block, and masked multi-head self-attention block.
Specifically, we replaced the masked multi-head self-attention block with an S4 block and removed positional encoding to compose our proposed model (see Fig.~\ref{fig:proposed}).
We utilized source-target attention to connect the encoder output and the S4 decoder in the same way as the Transformer decoder.
Each block contains a residual connection, dropout, and layer normalization~\cite{ba2016layer}.
We employed a linear layer and gated linear unit (GLU)~\cite{dauphin2017language-glu} activation for ensuring non-linearity after the S4 layer.
Unlike self-attention, S4 does not require positional encoding to handle positional information.
Therefore, we feed input vectors to S4 blocks without adding positional information.
During the training of S4 the decoder, we performed parallel computing using the convolution kernel of S4.
The prediction was executed in an autoregressive manner based on the RNN nature of S4.

In this paper, we applied the proposed S4 decoder to sequence-to-sequence ASR and TTS models.
In ASR, a text sequence was predicted in an autoregressive manner using the S4 decoder.
In TTS, the S4 decoder was used to generate acoustic features such as mel-spectrograms.

\section{Experiments}

\label{sec:experiments}
To verify the effectiveness of the proposed method, we evaluated its performance on ASR and TTS tasks. 
The proposed model was compared by simply replacing the self-attention of the baseline transformer decoder with the S4 layer.

\subsection{Automatic speech recognition}

\subsubsection{Experimental conditions}

We evaluated our proposed model on two corpora: CSJ~\cite{maekawa-etal-2000-spontaneous} and LibriSpeech~\cite{Panayotov2015-LibriSpeech}.
The CSJ corpus contains 581 h of Japanese speech sampled at 16 kHz and its transcription.
The LibriSpeech corpus contains 960 h of English speech sampled at 16 kHz for training an acoustic model and an additional 800M word token text-only corpus for building the language model.
For the evaluation on CSJ, we used the same evaluation sets as in ESPnet.
For the evaluation on LibriSpeech, we used the sets of dev-clean, dev-other, test-clean, and test-other. Each evaluation set contained 5 h data, and the ``other'' sets were more challenging to recognize than the ``clean'' ones.

For all the experiments, we used the ESPnet~\cite{watanabe2018espnet} toolkit for training and evaluation.
The basic configuration and preprocessing followed the LibriSpeech Conformer recipe\footnote{\url{https://github.com/espnet/espnet/blob/master/egs2/librispeech/asr1/conf/tuning/train_asr_conformer10_hop_length160.yaml}}, which consists of 12 encoder layers and 6 decoder layers.
We trained 50 epochs for the CSJ corpus and 60 epochs for the LibriSpeech corpus using the AdamW~\cite{loshchilov2017decoupled} optimizer with an exponential learning rate decay scheduler (40,000 steps for warmup, and a peak learning rate of 0.025).
Moreover, we excluded the weight decay from the embedding layer, normalization layer, bias parameters, and S4 parameters.
The numbers of dimensions in hidden layer and state space were 512 and 64, respectively.
We employed SpecAugment~\cite{park2019specaugment} and speed perturbation\cite{ko15_interspeech-speed} as data augmentation.
After finishing all the training epochs, we applied model averaging among the weights of the 10-best validation accuracy models during training. 
For the decoding process, we used beam search decoding.
The beam size was 25 for CSJ, 60 for LibriSpeech.
For each residual connection, we applied stochastic depth ($p = 0.1$) regularization~\cite{huang2016stochastic_depth}. 

\subsubsection{ASR results}
\begin{table}[t]
    \centering
    \caption{\it CER[\%] results on CSJ.
    The values except ``This work'' are those reported in other reference papers.
    \#Params(M) refers to the number of parameters in millions and $\dag$ refers to the result obtained using LM rescoring.}
    \vspace{2mm}
    \scalebox{0.85}[0.85]{
    \begin{tabular}{lcccc}
        \toprule
        Method          & \#Params(M) & \multicolumn{3}{c}{CER[\%]($\downarrow$)} \\
        \cmidrule(l{2mm}r{2mm}){3-5}
                      &   & eval1 & eval2 & eval3 \\
        \midrule
        \textbf{Hybrid} \\
        \quad Transformer~\cite{karita2019comparative}$\dag$ & - & 4.7 & 3.7 & 3.9  \\
        \quad Conformer~\cite{guo2021recent}$\dag$ & 91 & 4.5 & 3.3 & 3.6  \\
        \textbf{Transducer} \\
        \quad Conformer~\cite{Karita2021ACS} & 120 & 4.1 & 3.2 & 3.5 \\
        \midrule
        \textbf{This work (Hybrid)} \\
        \quad Transformer dec.       & 113.5 & 3.81	& 2.82 & 3.12 \\
        \quad S4 dec.                & 110.5 & \textbf{3.80} & \textbf{2.63} & \textbf{2.98} \\
        \bottomrule
    \end{tabular}
    }
    \label{tab:cer_on_csj}
    \vspace{-2mm}
\end{table}
Table~\ref{tab:cer_on_csj} shows the results of the CER on CSJ.
We confirmed that S4 decoder (S4 dec.) yielded lower CER values compared with the Transformer decoder (Transformer dec.) on all eval1/eval2/eval3 sets.
Table~\ref{tab:librispeech_result} shows the results of our model on the LibriSpeech dataset compared with the result of recently published models. 
In our experiment comparing S4 decoder with Transformer decoder, the WERs without the language model (LM) of S4 decoder were lower than those of Transformer decoder, and the WERs with the LM were comparable with those of Transformer decoder.
\begin{table*}[tbh]
    \centering
    \caption{\it WER[\%] results on LibriSpeech.
    The values except ``This work'' are those reported in the reference papers.
    \#Params(M) refers to the number of parameters in millions. In our experiments, we trained the Transformer language model followed by the ESPnet recipe and we obtained 30.88 of test perplexity.}
    \begin{center}
    \scalebox{0.85}[0.85]{
        \begin{tabular}{lcccccccccc}
        \toprule
        Method               & \#Params(M) & LM & \multicolumn{4}{c}{WER[\%]($\downarrow$) w/ LM} &
        \multicolumn{4}{c}{WER[\%]($\downarrow$) w/o LM}  \\
        \cmidrule(l{2mm}r{2mm}){4-7} \cmidrule(l{2mm}r{2mm}){8-11}
        & & & dev-clean & dev-other& test-clean& test-other& dev-clean& dev-other& test-clean& test-other \\
        \midrule
        \textbf{Hybrid} \\
        \quad Transformer~\cite{karita2019comparative} & - & Transformer & 2.2 & 5.6 & 2.7 & 5.7 & - & - & - & - \\
        \quad Conformer\footnotemark[2] & 116.2 & Transformer & 1.8 & 4.1 & 1.9 & 4.3 & 2.1 & 5.4 & 2.3 & 5.4 \\
        \quad SRU++~\cite{pan2022sru++} & - & Transformer & 1.9 & 4.8 & 2.0 & 4.7 & - & - & - & -  \\ 
        \quad Branchformer~\cite{peng2022branchformer} & 116.2 & Transformer & 1.9 & 4.2 & 2.1 & 4.5 & 2.2 & 5.5 & 2.4 & 5.5 \\
        \textbf{Transducer} \\
        \quad Transformer\cite{zhang2020transformer-transducer} & 139 & Transformer & - & - & 2.0 & 4.6 & - & - & 2.4 & 5.6\\
        \quad Conformer(L)~\cite{Gulati2020-conformer} & 118.8 & LSTM & - & - & 1.9 & 3.9 & - & - & 2.1 & 4.3 \\
        \quad ContextNet(L)~\cite{han2020contextnet} & 112.7 & LSTM & - & - & 1.9 & 4.1 & - & - & 2.1 & 4.6 \\
        \midrule
        \textbf{This work (Hybrid)} \\
        \quad Transformer dec. & 116.2 & Transformer & 1.81 & \textbf{3.98} & 1.95 & \textbf{4.21} & 2.18 & 5.50 & 2.43 & 5.53 \\
        \quad S4 dec. & 113.2 & Transformer & \textbf{1.72} & 4.10 & \textbf{1.88} & 4.25 & \textbf{2.07} & \textbf{5.31} & \textbf{2.29} & \textbf{5.13} \\
        \bottomrule
        \end{tabular}
        }
    \end{center}
    \label{tab:librispeech_result}
\end{table*}
\subsubsection{Robustness on long-form input}
We evaluated our model on the long-form utterance input situation because this also occurs in a real environment (\textit{e.g.}, difficult situations for voice activity detection (VAD), such as multi-speaker environment and noisy situations, etc.).
Pan \textit{et al.}~\cite{pan2022sru++} showed that the positional encoding layer in the Transformer-based model caused a lack of robustness in long-form input scenarios.
Further, we verified the robustness of S4 in long-form input scenarios by comparing it with the Transformer model.
We followed the experiment settings in~\cite{pan2022sru++}.
First, we prepared new evaluation datasets containing long-form audio.
We concatenated these consecutive three utterances in the CSJ/LibriSpeech corpora into one long-form speech utterance. 
Second, we compared each result of the CER/WER for the Transformer decoder and the S4 decoder on both the new evaluation datasets.

\begin{figure}[t]
  \begin{minipage}[b]{1.0\linewidth}
    \centering
    \includegraphics[clip,width=1.0\columnwidth]{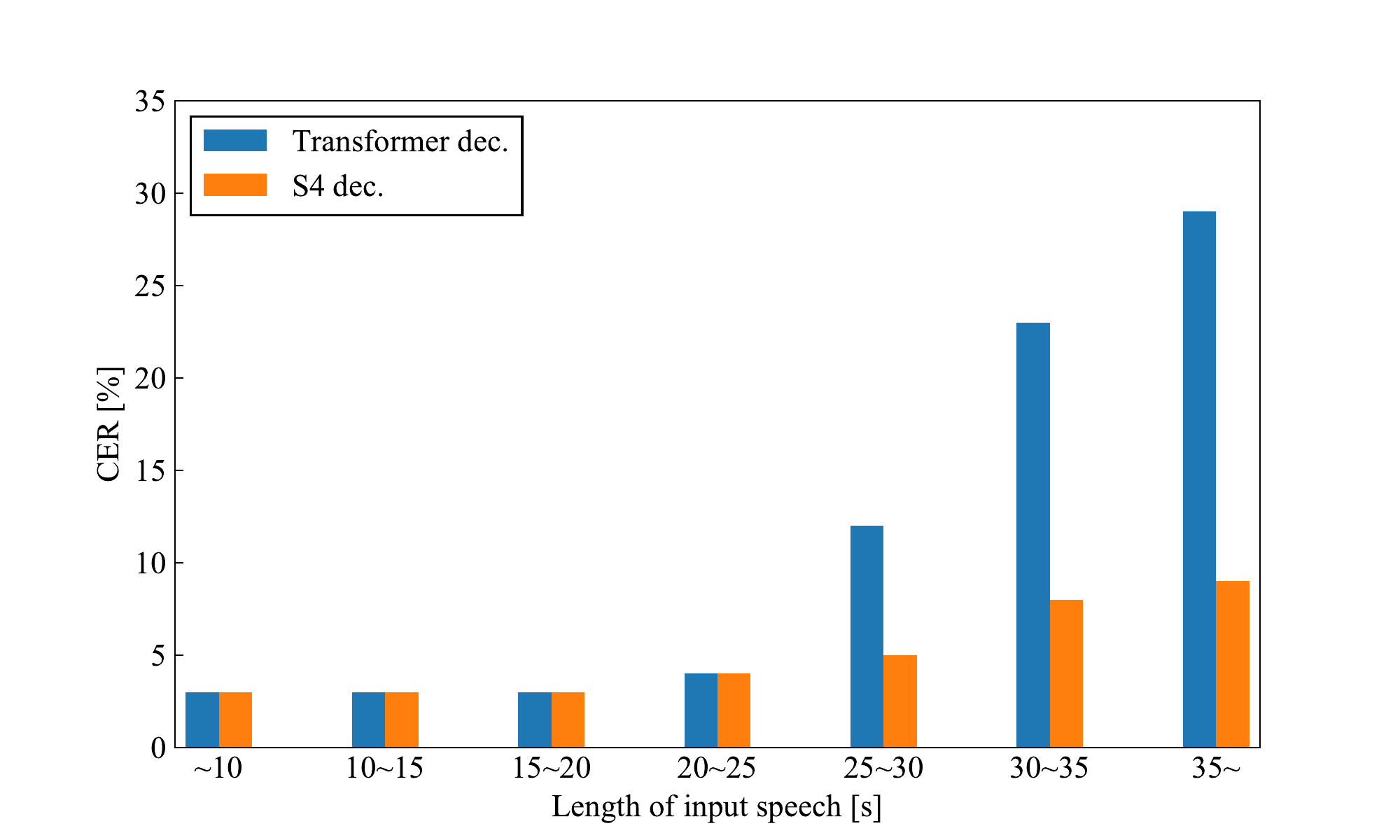}
    {(a) CSJ}
  \end{minipage}
  \begin{minipage}[b]{1.0\linewidth}
    \centering
    \includegraphics[clip,width=1.0\columnwidth]{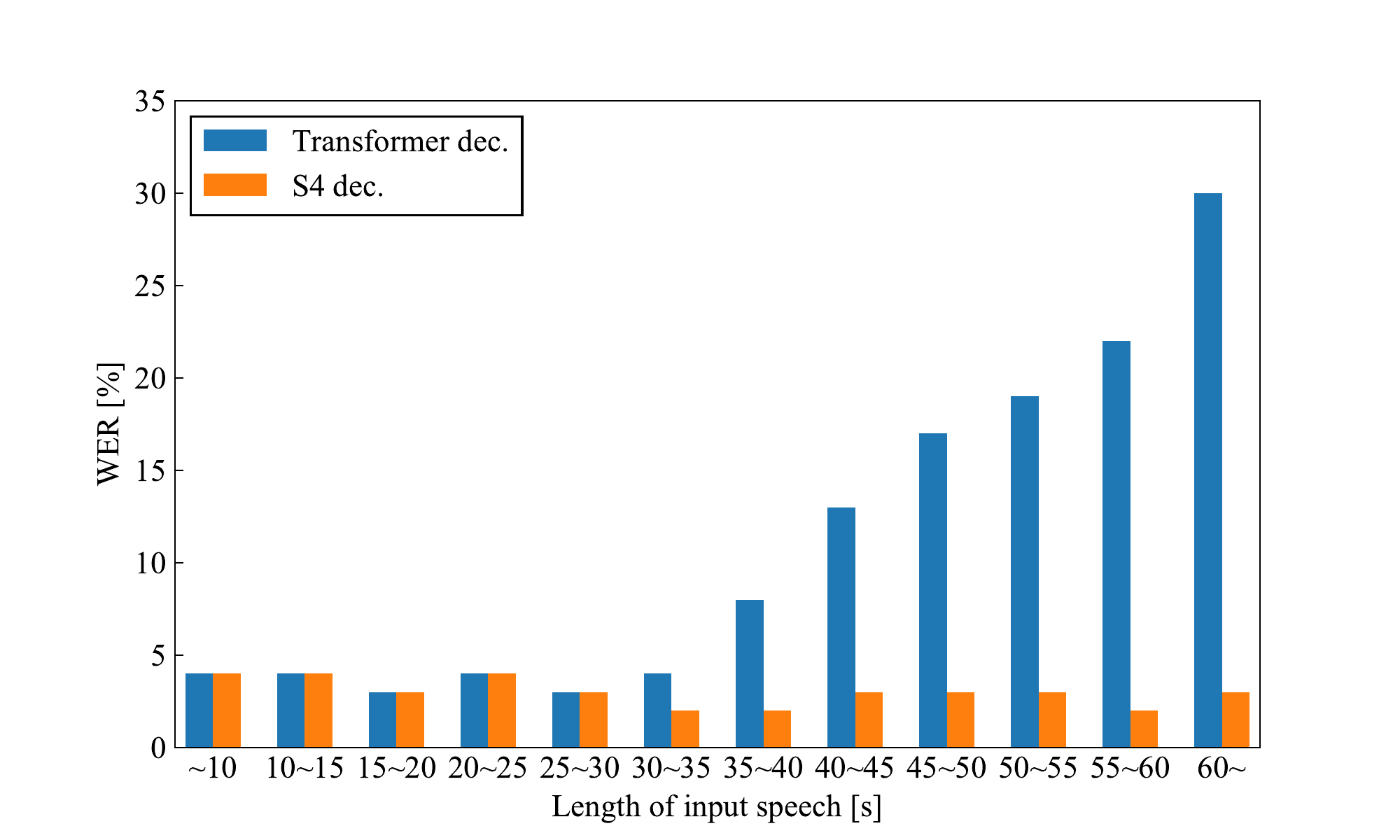}
    {(b) LibriSpeech}
  \end{minipage}
  \caption{\it{(a) CER on CSJ/(b) WER on LibriSpeech distribution by the speech length.}}
  \label{fig:long_audio}
\end{figure}

Fig.~\ref{fig:long_audio} shows the CER/WER distribution in terms of the speech length. We found that the S4 decoder was more robust than the Transformer decoder, particularly for audio longer than 30 s on both the datasets, consistent with the result reported in ~\cite{pan2022sru++}.
The results suggests that S4 better dealt with long-form sequences whose lengths were unseen during training than the Transformer model.
We speculated that the cause for the difference of error rates could be that the Transformer model had a non-trivial positional encoding layer; therefore, it could not deal with such a long-form audio.
However, the S4 model contained positional information implicitly in their model architecture, instead of positional encoding.

\subsection{Text-to-speech}
\subsubsection{Experimental conditions}

We evaluated the effectiveness of the S4 decoder using the TTS task. We used the ESPnet2-TTS~\cite{hayashi2021espnet2} framework with the LJSpeech~\cite{ljspeech17} dataset, containing 24 h of audiobook speech uttered by a single female speaker.
The dataset had 13,100 utterances, and we used 250 utterances each for the development and evaluation set.
For the proposed TTS task with the S4 decoder, we followed the Transformer recipe\footnote{\url{https://github.com/espnet/espnet/blob/master/egs2/ljspeech/tts1/conf/tuning/train_transformer.yaml}} by simply replacing the self-attention layer with a S4 layer in the Transformer decoder.
For the baseline methods, we used two autoregressive models (Tacotron2~\cite{shen2018natural} and Transformer-TTS~\cite{li2019neural}) and one non-autoregressive model (Conformer-FastSpeech2 (CFS2)~\cite{guo2021recent}).
We used pre-trained models available on \texttt{espnet\_model\_zoo}\footnote{\url{https://github.com/espnet/espnet_model_zoo}}.
Each model outputs mel-spectrograms as acoustic features.
\begin{figure}
    \centering
    \includegraphics[clip,width=1.0\columnwidth]{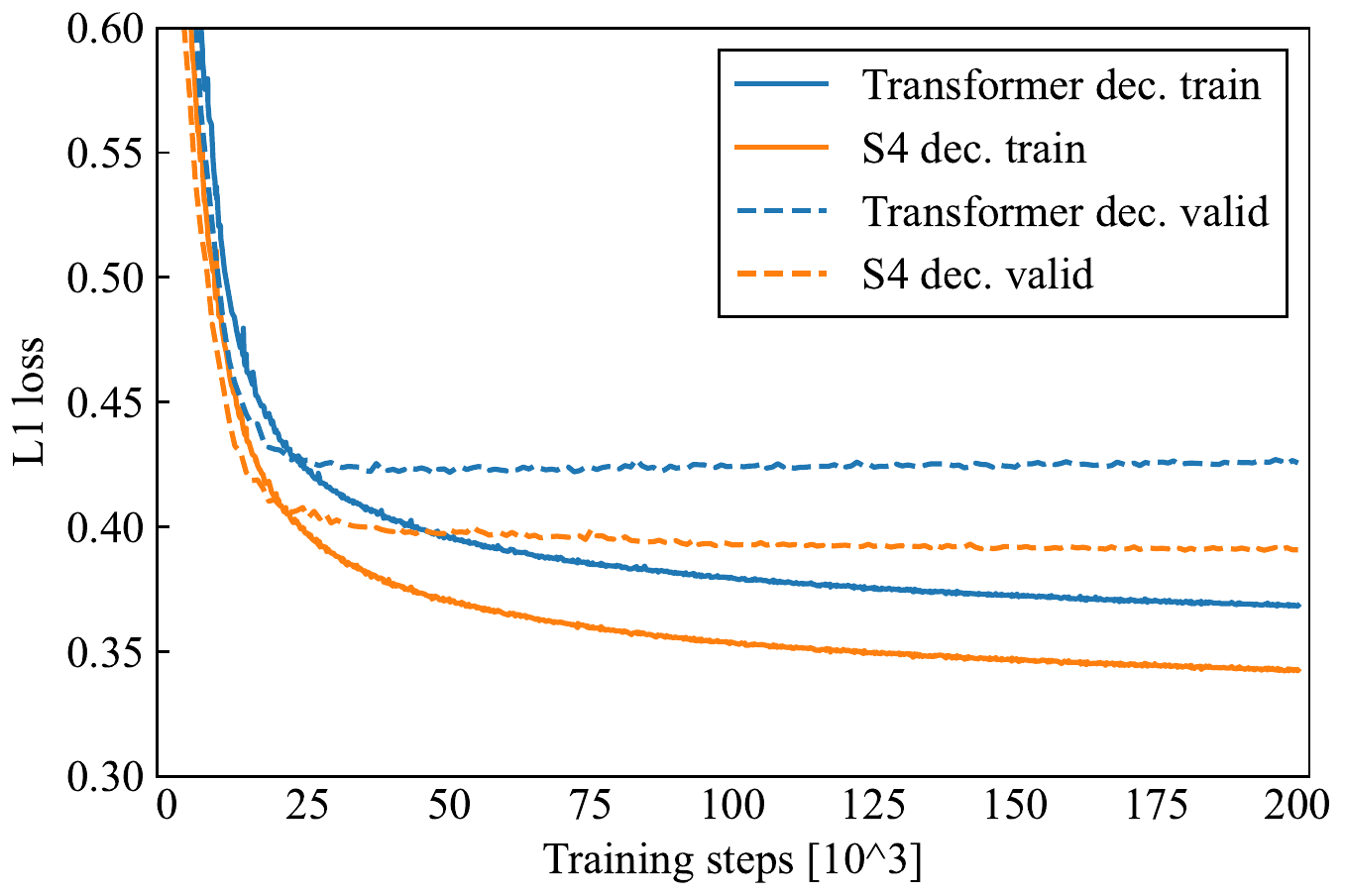}
    \caption{\it{L1 loss curves between target and generated acoustic feature.
    Training loss curves are plotted using solid line and validation loss curves are plotted using dashed line.}}
    \label{fig:loss}
    \vspace{-5mm}
\end{figure}
For the conversion from the generated mel-spectrogram to waveforms, we used the pre-trained HiFi-GAN vocoder~\cite{kong2020hifi}\footnote{\url{https://github.com/kan-bayashi/ParallelWaveGAN}} with the same split dataset.

\subsubsection{TTS results}
We observed loss curves of the proposed model with the S4 decoder.
Fig.~\ref{fig:loss} shows the L1 loss between the target and the generated acoustic features compared to the Transformer decoder.
The S4 decoder was found to converge to a lower value than Transformer decoder.
This result suggests that the S4 decoder has a potential for better generalization.

We evaluated the performance of the proposed TTS with the S4 decoder via objective and subjective evaluations.
For objective evaluation, we used the mel-cepstral distortion (MCD), log-$F_0$ root-mean-square error (log$F_0$RMSE) with the $F_0$ range restricted to [80, 400] (Hz), and CER.
To calculate the CER, we used a pre-trained ASR model\footnote{\url{https://zenodo.org/record/4037458}} and removed punctuation marks from the target text, and added 0.25 s of silence segments to the start and ending of the audio as a preprocessing.
Table~\ref{tab:tts_result} shows the results of the objective evaluation.
The experimental results show that the autoregressive model underperformed non-autoregressive models. This is because autoregressive TTS models suffered from attention errors, which caused misalignments between the input text and output acoustic feature.
The distortions and errors of autoregressive models will be mitigated by applying the dedicated methods proposed~\cite{he2019robust} to overcome such a misalignment problem.
However, we did not apply them in this experiment and will address them in the future.
Nevertheless, S4 decoder yielded lower MCD and log$F_0$RMSE than other autoregressive models.

Our subjective evaluation was conducted by a mean opinion score (MOS) test on Amazon Mechanical Turk.
The number of participants was 50 and each participant listened to 30 speech samples composed of five randomly chosen sentences with six methods including recorded and vocoder reconstructed samples.
Participants rated the naturalness of samples on a five-point scale (1 = bad, 2 = poor, 3 = fair, 4 = good, and 5 = excellent).
Table~\ref{tab:tts_result} shows the MOS result. 
Our proposed S4 decoder yielded higher MOS than the Transformer, which indicated better acoustic feature generation than the self-attention.
Furthermore, despite the existence of a misalignment, the MOS of the S4 decoder outperformed that of CFS2.

\begin{table}[t]
    \centering
    \caption{\it TTS results. GT refers to the recording sample, GT (mel) refers to a reconstructed sample with vocoder, and CI refers to 95 \% confidence interval}
    \vspace{2mm}
    \scalebox{0.80}[0.80]{
    \begin{tabular}{lcccc}
        \toprule
        Method          & MCD[dB]($\downarrow$) & log$F_0$RMSE($\downarrow$) & CER[\%]($\downarrow$) & MOS($\uparrow$)$\pm$CI \\
        \midrule
        GT              &  N/A  &   N/A & 1.0 &      4.33 $\pm$ 0.10    \\
        GT(mel)         &  2.64 & 0.110 & 1.1 &      4.08 $\pm$ 0.11       \\
        \midrule
        Tacotron2       & 7.18 & 0.280 & 2.0 &      3.47 $\pm$ 0.13 \\
        Transformer     & 7.02 & 0.255 & 3.5 &      3.74 $\pm$ 0.12   \\
        CFS2            & \textbf{6.46} & \textbf{0.227} & \textbf{1.2} & 3.70 $\pm$ 0.13\\
        S4 dec. (ours)      & 6.87 & 0.243 & 2.7 & \textbf{3.92 $\pm$ 0.12} \\
        \bottomrule
    \end{tabular}
    }
    \vspace{-2mm}
    \label{tab:tts_result}
\end{table}

\section{Conclusion}
\label{sec:conclusion}
In this study, we evaluated the effectiveness of the S4 decoder on ASR and TTS tasks. The S4 decoder produced a comparable performance with the Transformer decoder.
We found that our S4 decoder could handle long-form sequence inputs without performance degradation. Moreover, our S4 decoder achieved better generalization performance than the Transformer decoder on a TTS task.
Therefore, we believe that the S4 decoder has good application potential in various autoregressive models and tasks that require them.
In the future, we plan to investigate the effects of using S4 as a transducer model and apply it to the encoder part. 
Furthermore, we aim to identify the relationship between lack of robustness and positional encoding layer in ASR tasks.

\newpage
\newpage
\clearpage

\section{References}
\printbibliography


\end{document}